\documentclass[psf,epsf,twocolumn,showpacs,preprintnumbers]{revtex4}
\usepackage{graphics}
\usepackage{graphicx}
\usepackage{dcolumn} 
\usepackage{bm}
\usepackage{epsfig}
\begin{document}
\title{Boron Spectral Density
      and Disorder Broadening in B-doped Diamond}
\author{K.-W. Lee and W. E. Pickett}
\affiliation{Department of Physics, University of California, Davis, CA 95616}
\date{\today}
\begin{abstract}
Comparison of periodic B dopants with a random alloy of substitional
boron in diamond is carried out using several supercells and the coherent
potential approximation (CPA) for the random alloy case.  
The main peak in the B local density of states is shifted 
to lower binding energy compared to the corresponding 
C peak in intrinsic diamond.
In supercells, this shows up as strongly B-character bands split
from bulk C bands away from the zone center,
in an energy region around -1 eV.  
Even for a 4$\times$4$\times$4 supercell (BC$_{127}$), effects of the
dopant order are evident in the form of primarily B-character bands
just below the Fermi level at the supercell zone boundary.
The bands resulting from the CPA are of continuous mixed C-B character.  They 
resemble virtual crystal bands,
but broadened somewhat reflecting the disorder-induced lifetime, 
and are consistent with angle-resolved photoemission band maps.
The B character is 1.7 times larger than for  C (per atom)
near the top of the valence bands for CPA, and roughly the same 
for supercells.
CPA results are particularly useful since they characterize
the wavevector and energy dependence of disorder broadening.

\end{abstract}
\maketitle

\section{introduction}
The discovery and confirmation of superconductivity in diamond
heavily doped with 
boron\cite{ekimov1,takano,bustarret,ekimov2,umezawa,takano2} 
has provoked interest in the
mechanism and character of the pairing that is responsible.  
There are several confirmations of superconductivity around 4 K
for 2-3\% doping,  and a recent report\cite{umezawa} indicates
T$_c \approx$ 11.5 K for 4.7\% boron concentration in 
homoepitaxial films.
While
the picture of 3\% hole-doped diamond as a degenerate metal has
been the most common 
assumption,\cite{ucd1,boeri,blase,xiang,ma,cardona} 
there has also been the
possibility that the boron acceptor level is too deep for 
itinerant conductivity to arise, leading to a picture of a
non-degenerate doped semiconductor\cite{pogorelov} with possible
correlated behavior.\cite{baskaran} 
An angle-resolved photoemission spectroscopy (ARPES) study\cite{arpes} 
on heavily B-doped
diamond has found the Fermi level to lie a few tenths of an eV within
the valence bands, establishing the degenerate-metal picture as the
correct one for this system.

It is useful to review briefly the history of B doping of diamond.
Although it has been clear for decades that B is responsible for 
the color of blue diamonds, and the desire for p-type semiconducting
diamond has been intense for some time, the existing information 
about the isolated B acceptor is sketchy.  The energy of the acceptor level
is agreed to be E$_B$=0.37 eV.\cite{thonke}  
This binding energy is too large for
this impurity to be treated accurately at the effective mass level,
hence its wave function is not as greatly extended as that of a shallow
level would be.   On the other hand, this is not truly a deep
level (since E$_B$ is only 7\% of the band gap) in which the defect 
state extends over only a shell or two
of neighboring C atoms. 

The existing theoretical studies aimed at gleaning information about the
B acceptor state are not very conclusive.   
Finite clusters (up to 59 atoms) were used by Mainwood,\cite{mainwood}
who obtained for substitutional B a threefold degenerate, mainly B
$2p$, state 0.36 eV above the uppermost occupied diamond valence state.
The cluster was allowed to relax, with the B-C bonds increasing by
10\% (0.15~\AA), a much larger amount than has been obtained in more 
recent calculations.  The very limited cluster size, with its surface
effects, does not allow for clear conclusions. 
Barnard {\it et al.}
reported some results from 64 atoms (periodic) supercells,\cite{barnard} 
but as we show here
for a larger supercell, the local density approximation (LDA) does
not produce a gap acceptor state at such densities ($\sim$ 1\%).
A tight-binding total energy method was used by Sitch {\it et al.},
who reported only atomic relaxation using a 216 atom 
supercell.\cite{sitch}
C atoms neighboring the substitutional B relaxed outward by 4\%
(bond length of 1.60~\AA).  Their study was focused instead on doping
of amorphous carbon.

Adopting the ARPES demonstration\cite{arpes} that the Fermi level 
lies within the itinerant valence bands,
there are a few choices to be made in describing the electronic structure,
and the subsequent evaluation of the character and strength of the
electron-phonon (EP) coupling is determined to some extent by this choice.
The published studies of superconductivity in B-doped diamond concur
that EP coupling provide the pairing mechanism, in spite
of applying different approximations to achieve their separate estimates.
We briefly describe the various approaches.

Three groups have adopted the virtual crystal approximation (VCA), making use of
the fact that B is chemically and physically similar to C, especially 
when they share a common environment; in this case each is surrounded by
C atoms on a diamond lattice.  In such cases, the 
itinerant wavefunctions average
over the B and C atoms and replacing each nucleus by an average 
charge constitutes the VCA.   Adopting the VCA, Boeri, Kortus, and
Andersen\cite{boeri} then applied linear response theory for harmonic
phonons and calculated the phonon frequencies at $\sim$50 q points.  This 
allowed the evaluation of the EP spectral function 
and strength
of coupling.  Lee and Pickett\cite{ucd1} also adopted
the VCA for the underlying electronic structure.  As suggested by the 
analogy with MgB$_2$ where holes are doped into strong covalent bonds,
they  found a very large
EP matrix element for the bond-stretch mode and, making use of the
small Fermi surfaces, assumed it varied slowly
with wavevector q up to 2k$_F$ (k$_F$ is the Fermi wavevector).  
They calculated by the frozen phonon
method the phonon renormalization of the bond-stretch mode
by the doped carriers, and from two
related but separate directions arrived at the EP coupling strength.
Also adopting the VCA viewpoint, Ma {\it et al.}\cite{ma} used methods
very similar to those of Boeri {\it et al.} (but different codes) and the
same phonon mesh, but studied electron doping as well as hole doping.

Blase, Adessi, and Conn\'etable\cite{blase} and Xiang {\it et al.}\cite{xiang},
on the other hand, used a supercell approach in which the B dopants 
keep their chemical identity (actual B atoms), but to apply standard
solid state methods the dopant atoms had to be 
repeated periodically.  This supercell approach
restricts dopant concentrations to certain
commensurate fractions but this is no real limitation because B 
concentrations are only known approximately.  Blase {\it et al.} calculated
phonons and EP coupling only at q=0 in a 54 atom (BC$_{53}$) supercell and
estimated the coupling strength accordingly.  Xiang {\it et al.} treated
both 16 atom and 36 atoms supercells, using phonons at a few q points. 

Naturally the various calculations contain differences in detail and in
general it is not possible to compare several of the
intermediate results.
Both groups using the supercell approach have suggested that there are
aspects of their results that make the supercell approach superior, or
perhaps necessary.  On the other hand, the ARPES data\cite{arpes} lead to band
dispersions very much like the diamond (or VCA) bands, perhaps broadened
by disorder.

There is another approach to the treatment of randomly positioned
dopant atoms, the coherent potential approximation (CPA).  The CPA
allows the dopant to retain its identity, but produces an effective
medium that accounts for random disorder.  The CPA method has not been
applied to evaluate the EP coupling in B-doped diamond, which is possible
in principle, because linear response theory for phonons has not yet
been incorporated into CPA codes.  It is straightforward however to
calculate the CPA electronic structure of B-doped diamond and compare
results with the VCA and supercell approaches. 
In this paper we study in Sec. III selected supercells, and then provide
CPA results (Sec. IV) for
dilute B in diamond and make some comparisons both to supercell results
and the VCA bands (which are hardly distinguishable from the diamond bands
for small B concentrations).  Using supercells 
we can study effects due to
B-B interaction, and quantify specific effects that arise because of the
periodicity that is imposed.  It is also easy when using the supercell
approach to evaluate the
effect of B on the various shells of C neighbors.  The CPA accounts for
random  positioning of the dopants and also allows
the evaluation of disorder broadening of the bands (Sec. IV).
We make a few observations about relaxation around a substitutional B
atoms (Sec. V) which is not included in either the VCA or CPA method, 
and in Sec. VI provide a summary of the implications of
our results.

\section{Structure and calculation method}
\label{structure}
 The bulk diamond bond length is d$_{\tt C-C}$= 1.54 \AA.
 For a substitutional B atom, the B$-$C bond length is 
 increased by 2\%
 and the neighboring C-C bond length is decreased by 1.3\% in a
 LDA calculation with 
 $3\times3\times3$ fcc primitive cell (BC$_{53}$).\cite{blase}

 We have carried out calculations of ``ordered impurities" using 
several supercells and of ``randomly disordered impurities" 
using the CPA method, both of them 
with the full-potential nonorthogonal 
local-orbital minimum-basis scheme (FPLO).\cite{klaus1} 
The CPA\cite{klaus2} method implemented in FPLO
is based on the Blackman-Esterling-Berk theory\cite{blackman} that
includes random off-diagonal matrix elements in the local orbital
representation.  Koepernik {\it et al.} provide a discussion of 
how randomness is treated in this atomic-orbital
based procedure.\cite{klaus2}

We have used three types of supercells: 
a large symmetric cell, a doubled cubic cell only in
a (001) plane and a set of layered cells.
A $4\times4\times4$ fcc primitive cell (BC$_{127}$) was used
as a symmetric cell and $2\times2\times1$ cubic cell (BC$_{31}$)
also was studied.
The latter ($\approx 3\%$ doping) is very close to the initally reported 
$2.5\%$ doping superconductor.
In addition, we used several supercells 
[($1\times1\times2$ (BC$_{15}$) to $1\times1\times5$ (BC$_{39}$)]
layered in the $c$-direction, which illustrated some systematic
variations.
The low level of B doping requires a sufficiently fine $k$ point
grid to sample the few unoccupied valence states that arise.
Because of the different size of each Brillouin zone (BZ), $k$ points
used are different from each other, but we have used from 245 to 1000 $k$
points in the irreducible wedge of each BZ,
except for $4\times4\times4$ supercell. For the 
$4\times4\times4$ supercell, 3 irreducible $k$ points should be 
enough to understand the band structure. 
However, CPA requires a much finer $k$ mesh to sample the 
wavevector and energy dependent spectral function, so 2045 $k$ points were used 
in the irreducible wedge.
For all calculations, valence orbitals for the basis set
contained $1s2s2p3d$ for B and C.

\section{Ordered Impurities (Supercells)}

\begin{figure}[tbp]
\rotatebox{-90}{\resizebox{5cm}{8cm}{\includegraphics{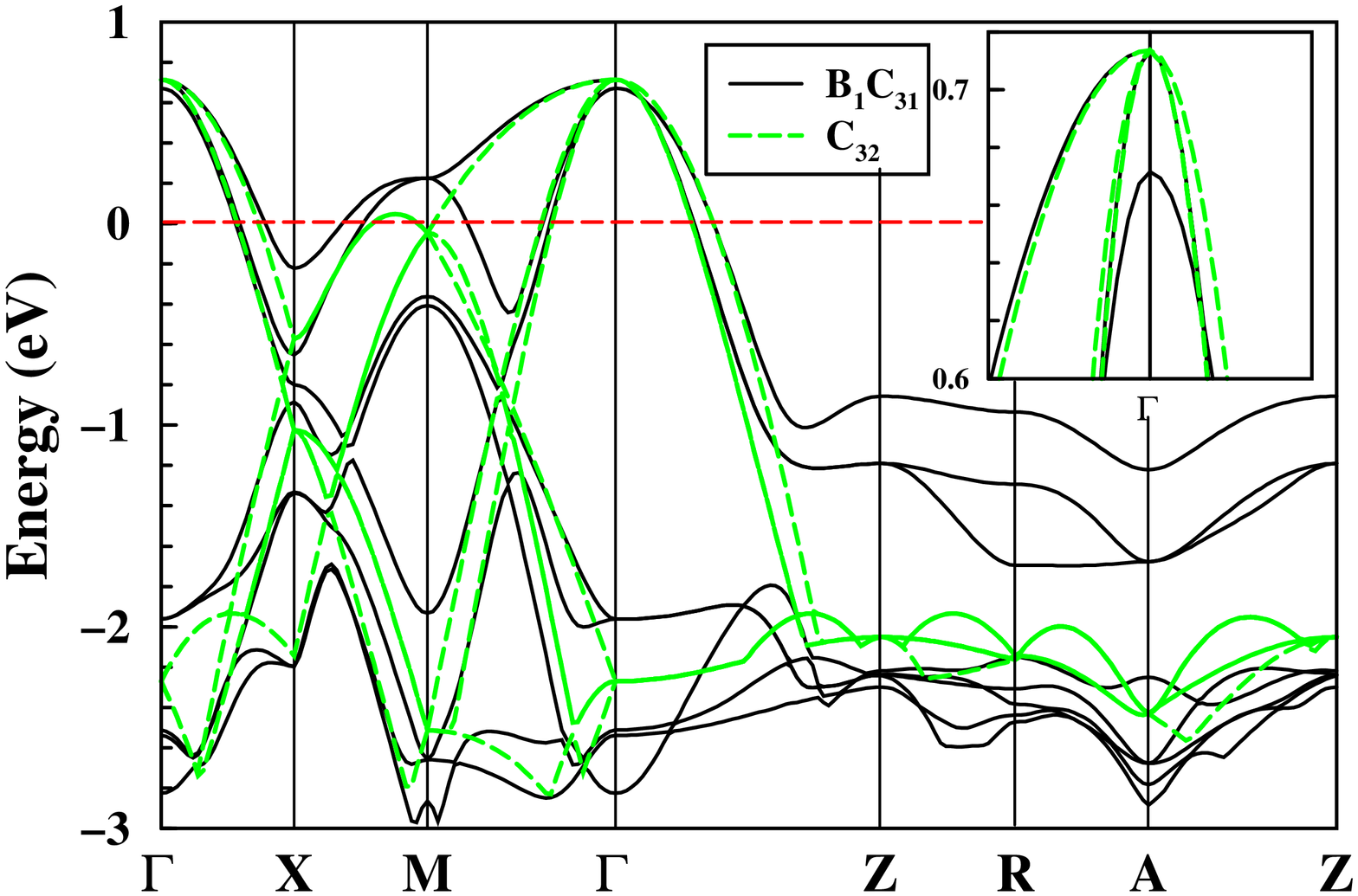}}}
\vskip 8mm
{\resizebox{8cm}{5cm}{\includegraphics{Fig1b.eps}}}
\caption{(Color online) {\it Top panel}: Upper portion of 
 the valence band structures 
 of the $2\times 2\times 1$ supercell for the B doped (BC$_{31}$) and 
 intrinsic diamond (C$_{32}$). 
 The bands are aligned at the top of the valence band.
 {\it inset}: Blowup of the band splitting by 42 meV along the M-$\Gamma$-Z
 lines due to the asymmetry of the supercell.
 {\it Bottom panel}: ``Fatband" representation of the B $2p$ character.
 The main character is located in the range of $-1.7$ eV and $-0.7$ eV,
 a regime largely shifted by the B-doping, on the $k_z=\pi/a$ plane.
 The symbol size is proportional to the B $2p$ character.
 The dashed horizontal line indicates the Fermi energy of BC$_{31}$.
 The symmetry points are given such as (0,0,$\xi$) for $\Gamma$(Z),
 (0,1/2,$\xi$) for X(R) and (1/2,1/2,$\xi$) for M(A). $\xi$ is zero 
 for the first symbols and 1 for the symbols in parentheses; units
 are $\frac{\pi}{a}$.}
\label{band221}
\end{figure}

\begin{figure}[tbp]
\rotatebox{-90}{\resizebox{7cm}{8cm}{\includegraphics{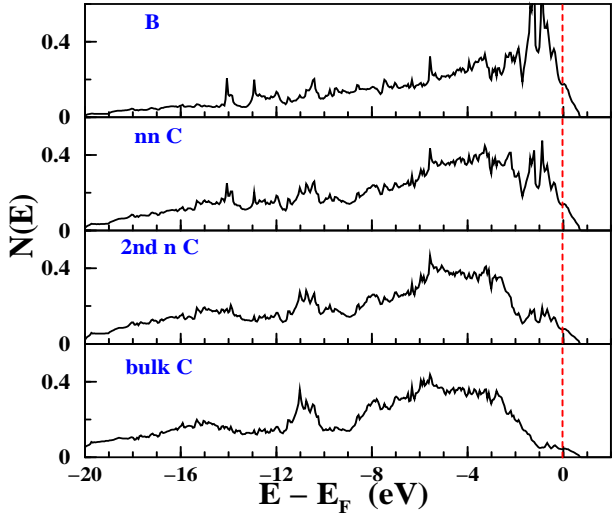}}}
\caption{(Color online) Atom-projected DOS for B and and three C neighbors
 in the $2\times 2\times 1$ supercell (BC$_{31}$).  The
 B DOS is very similar to that of the nn C.
 Note the sharp peaks confined to B and the nn C atom
 between $-2$ eV and $-1$ eV, arising
 from strong B $2p$ character on the plane of $k_z=\pi/a$.
 For C atoms progressively further from the B atom, 
 the peak at $-11$ eV 
 becomes sharper.  The B DOS is clearly shifted higher (to lower binding 
 energy) compared to that of carbon. }
\label{dos221}
\end{figure}

\subsection{$2\times2\times1$ (32 atom) supercell}
First we address the supercell having a B concentration
($c_B\sim 3\%$) very close to the initially reported superconducting
concentration ($\sim 2.5\%$).
The upper portion of the valence band structures of the 
$2\times 2\times 1$ supercell
is compared with that of intrinsic diamond in the top panel of 
Fig. \ref{band221}.
For better comparison, the bands of intrinsic diamond are raised by
the Fermi energy (E$_F$) difference from the doped diamond so that
the bands are aligned at the valence band maximum (VBM). 
B doping has three separate effects that are common to all of the
supercells: (1) lowering E$_F$ owing to hole doping (opposite direction
to the effect of C doping in Mg(B$_{1-x}$C$_x$)$_2$,\cite{ucd2}) 
(2) band splittings at the high symmetry points in the BZ 
reflecting symmetry breaking, and (3) bandshifts reflecting the 
difference in the B and C potentials.

The largest effect of the B doping occurs $1-2$ eV below the Fermi level,
which is easily seen on the $k_z=\pi/a$ plane 
(Z-R-A-Z lines in Fig. \ref{band221}). 
Three bands along these lines are raised in energy by more than 1 eV 
by the addition of boron;
the strong B character of these displaced bands is illustrated in
the bottom panel of Fig. \ref{band221}, where the ``fatband'' representation
emphasizing B character is presented.
These rather flat bands are strongly hybridized 
with the nearest neighbor (nn) C $2p$
states, which provide about one-third of the band character.
At the VBM which is threefold degenerate in intrinsic diamond,
the B character and the nn C $2p$ character are also strongly mixed.
Besides the mixing, the top of the valence bands at the $\Gamma$ point
in the B-doped diamond shows a tiny band splitting of 42 meV, as 
shown in inset of the top panel of Fig. \ref{band221}.
The origin of the splitting is not clear but
its value is probably specific to this supercell.

The corresponding atom-projected DOS are given in Fig. \ref{dos221}.
As expected from the strong local B$-$C mixing, the B DOS resembles that of the 
nn C DOS. 
More specifically, however, the B and nn C are enhanced at the VBM.
Their spectral densities show differences: the B DOS 
has much sharper peaks around $-1$ eV while that of the nn C  continues
more strongly to -4 to -6 eV.
To summarize: the weights of the DOSs are arranged
in the order of B, nn C, second neighbor (2nd n) C,
and bulk C, in order of increasing binding energy.\cite{2nnC}
This trend is consistent with the interpretation of x-ray absorption
and emission spectroscopy by Nakamura et al.\cite{nakamura,nakamura2}

\begin{figure}[tbp]
\vskip 8mm
\resizebox{8cm}{5cm}{\includegraphics{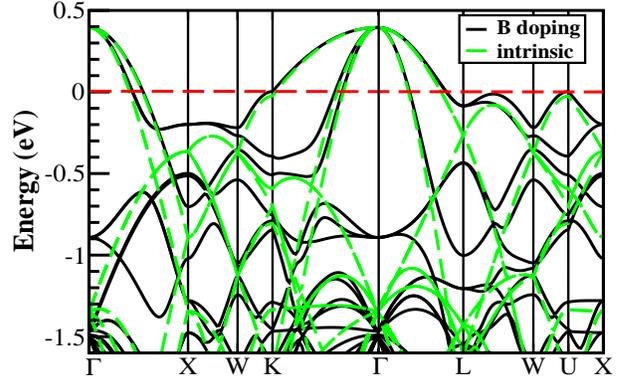}}
\caption{(Color online) Upper portion of the valence band structures of 
  $4\times4\times4$ supercell for the B doped (BC$_{127}$) and 
  intrinsic diamond (C$_{128}$).
  At the symmetry points, the band splittings are obvious, especially
  at the $\Gamma$, X, and L points.
  The symmetry points follow the conventional notation for
  fcc BZ. (L denotes the zone boundary point in $<111>$ direction.)
  The dashed horizontal line indicates E$_F$ of the B-doped diamond.}
\label{band444}
\end{figure}

\subsection{$4\times4\times4$ (128 atom) supercell}
As large a supercell calculation as possible is desirable to extract
information about the isolated B impurity.
We first study the change in the band structure, keeping in mind that
it reflects the periodic arrangement of the 0.8\% concentration of
B substitutionals in this supercell.  The symmetry is such that the
threefold degeneracy of the VBM is not split.  The effective masses are
changed very little, but we have not tried to identify possible changes
in masses at the 1\% level.  Splitting of degeneracies by $0.3-0.4$ eV
can be seen at various places in the bands in Fig. \ref{band444}: 
along $\Gamma$-X and at L near $-0.2$ eV. 
There are larger splittings (at L, for example) at energies 
below $-1$ eV in Fig. \ref{band444}.
At the $\Gamma$ point, the cluster of degenerate bands at $-1.3$ eV
in intrinsic diamond (coming from zone folding)
are split and 
shifted considerably by B addition.
The threefold degenerate bands at -0.9 eV, having strong B
character, have been raised by 0.5 eV, while others are
shifted little.
Thus even this large supercell shows effects of the periodicity
of B atoms that will not be present if B atoms are distributed randomly.
These effects are most evident when they
produce new pieces of Fermi surface related specifically to the
artificial periodicity, as is beginning
to be the case in Fig. 3 along the L-W line.  A higher doping level
than the 0.8\% of this supercell could lead to inaccuracies
in properties determined by the Fermi surface, such as the 
electron-phonon spectral function.

\subsection{Layered supercells}
\begin{figure}[tbp]
\rotatebox{-90}{\resizebox{5cm}{8cm}{\includegraphics{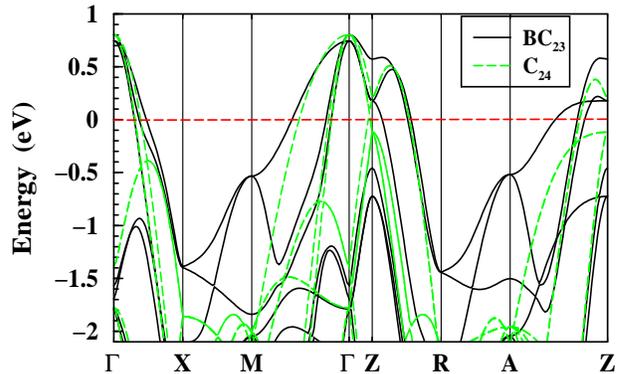}}}
\caption{(Color online) Upper portion of the valence band structures of the
 $1\times1\times3$ supercell for the B doped (BC$_{23}$) and
 intrinsic diamond (C$_{24}$). 
 This figure shows clearly the different origins of the band splittings
 at $\Gamma$ and at Z. 
 The dashed horizontal line indicates $E_F$ of the B-doped diamond.}
\label{113band}
\end{figure}

 A set of layered supercells were studied to observe closely 
the asymmetry, changes in doping concentration, 
and effect of increasing the separation
of the B atoms. 
Although the doped B concentration
(2.5\% to 6.25 \%) varies in the each calculation, 
the band structures show considerable similarities.
The upper portion of the valence band structures of the $1\times1\times3$
supercell is shown in Fig. \ref{113band}.
Bands of strong B $2p$ character are pushed upward into the -1.5 to -0.5 eV 
range along the X-M and R-A lines. 
This is the same energy range where B character is `pushed' in the
other supercells.

\begin{figure}[tbp]
\rotatebox{-90}{\resizebox{7cm}{7cm}{\includegraphics{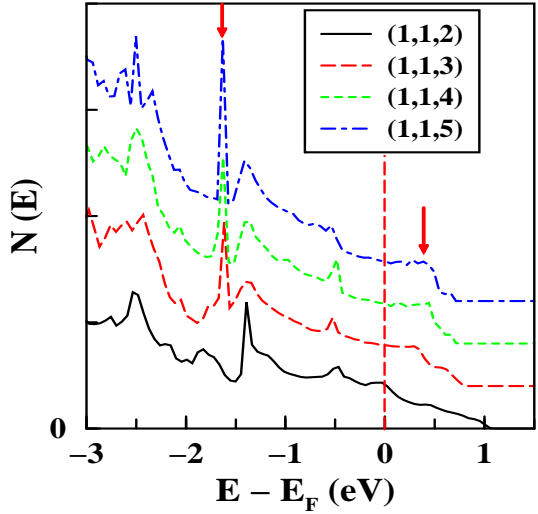}}}
\caption{(Color online) Total DOS of the (1,1,$m$) cube supercells 
 ($m$=2,3,4,5); in each case there is one hole per supercell unoccupied. 
 The arrows denote peaks with strong B character. 
 The Fermi energies of every plot are set to zero and spacing of
 each plot in $y$-axis is chosen as 2 states/eV.}
\label{zdos}
\end{figure}

The total DOS of the (1,1,$m$) cubic supercells ($m$=2,3,4,5),
corresponding to $x$=0.0625, 0.0417, 0.0313 and 0.025,
are presented in Fig. \ref{zdos} and show an interesting trend.
As indicated by the arrows, decreasing $x$ leads to a sharp peak at
$-1.6$ eV and enhancement around 0.4 eV.
With decreasing $x$, the lowest of the bands shown
in Fig. \ref{113band} along the X-M line becomes flatter,
resulting in enlargement of the peak at $-1.6$ eV.
As suggested by Xiang et al.,\cite{xiang} we find that the width of the acceptor
states (unoccupied portion of the valence bands), containing one hole
per supercell, depends on $x$ up to a certain value.
Note that the unoccupied part of N(E) is essentially the same in the
(1,1,4) and (1,1,5) supercells.

\section{Disordered boron (CPA)}
\begin{figure}[tbp]
\resizebox{7.5cm}{6.2cm}{\includegraphics{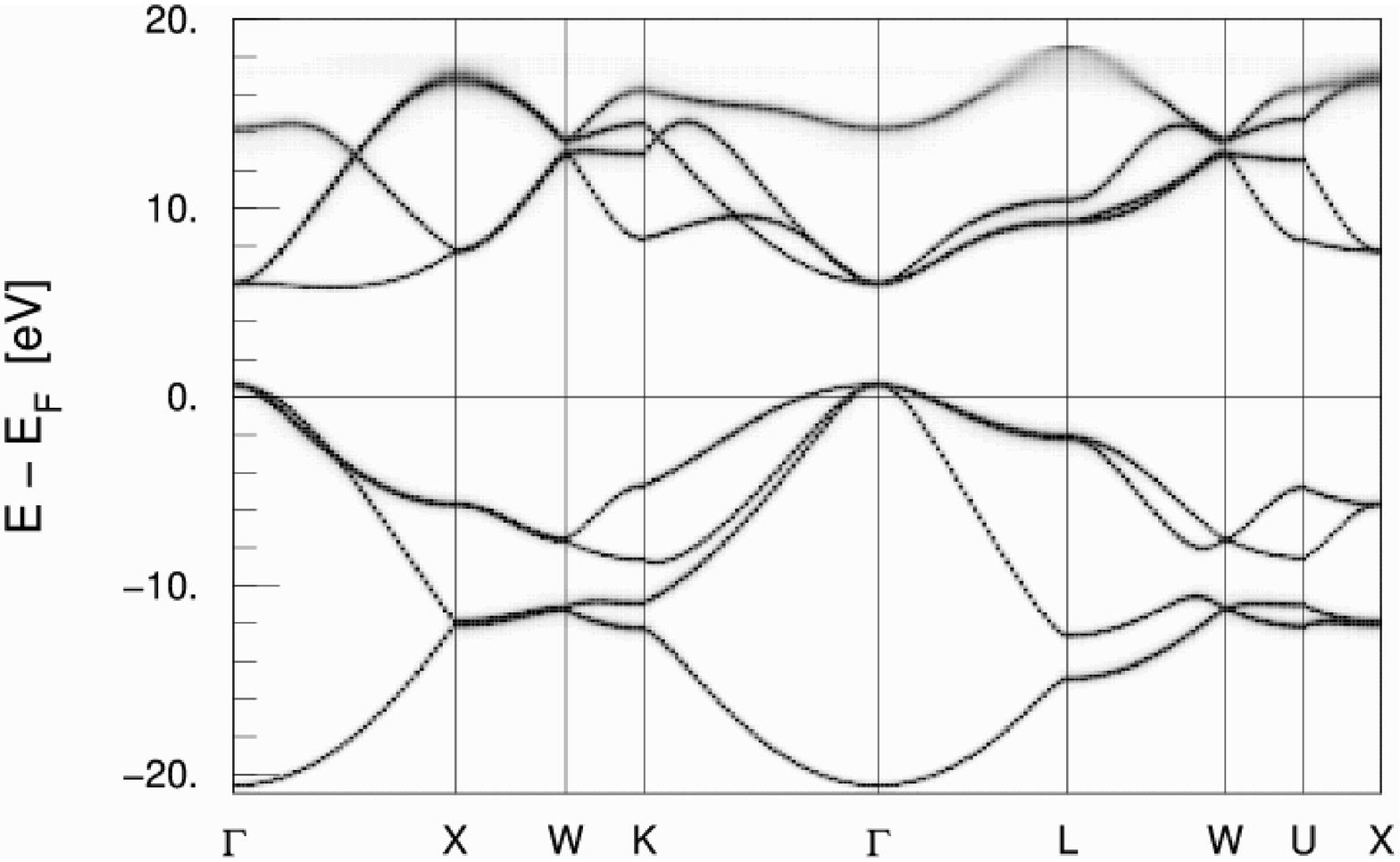}}
\caption{CPA spectral density, as a ``smeared" band structure,
  in the full valence-conduction band region for $x$=0.025.
  The conduction bands around 17 eV are most strongly affected
  by the chemical disorder, while the valence bands are less affected.}
\label{fullband}
\end{figure}

\begin{figure}[tbp]
\resizebox{8cm}{7cm}{\includegraphics{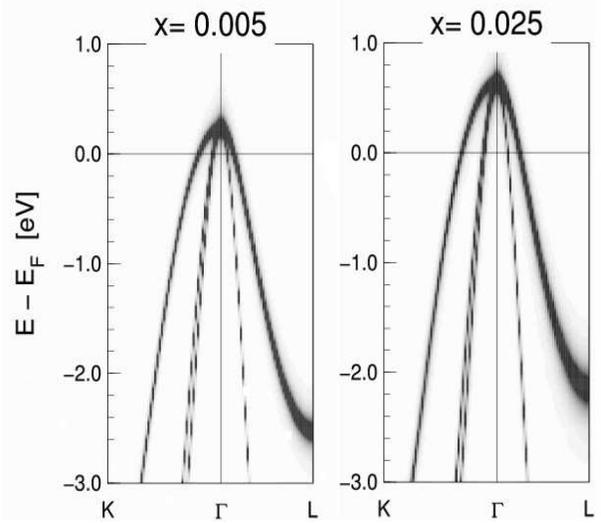}}
\caption{Blowup CPA spectral density, with Fermi levels aligned,
near $E_F$ along
the K-$\Gamma$-L line
for $x$=0.005 and $x$=0.025.
As $x$ increases, E$_F$ is lowered and the disordered
broadening increases. The spectral density shows the broadening most clearly
at the $\Gamma$ and L points.}
\label{smallband}
\end{figure}

\begin{figure}[tbp]
\rotatebox{-90}{\resizebox{5cm}{8cm}{\includegraphics{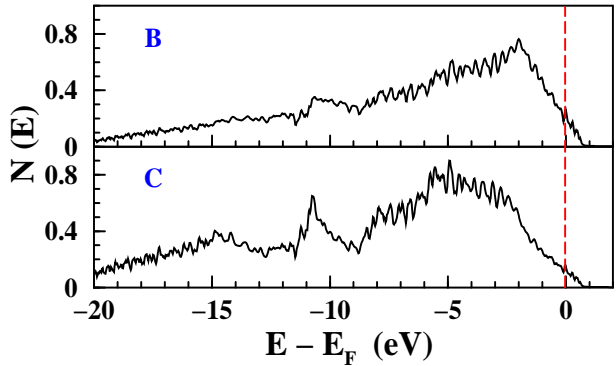}}}
\caption{(Color online) Atom-projected DOS at $x$=0.025 from 
 the CPA calculation. The picture that results is very similar 
 to that from the supercell calculations.
 The main peak of B is located at -2 eV,
 while that of C lies at higher binding energy and
 is almost identical to that of the bulk C. }
\label{cpados}
\end{figure}

Now we describe CPA calculations which explicitly treat randomly 
distributed B substitutionals with concentration $x$.
We calculate $x=$0.005=1/200  to approximate an isolated 
boron, and $x=$0.025 very close to the optimal superconducting
concentration $c_B$.
The CPA spectral density for $x$=0.025, rendered as a 
``smeared" band structure, 
is displayed
in the full band region in Fig. \ref{fullband}.
The bands are like those from the virtual crystal approximation
used earlier\cite{boeri,ucd1} except that lifetime effects (appearing
as a broadening) are visible.
The effect of disorder is strongest in the conduction (unoccupied) 
band near 17 eV. The valence (occupied) band is less affected. 
Even for only 2.5\% B-doping, the disordered broading is
clearly visible in the doubly degenerate band near $E_F$ along 
$\Gamma$-L, and near $-12$ eV along the X-W line. The former comes from
mixing with B $2p$ states and the latter has more admixture of B $2s$ states.
It is of course no accident that the disorder broadening resulting from
the CPA treatment is largest in the valence band (in the zero to -2 eV
region relative to E$_F$) where the B-derived
band shifts are largest.  The broadening arises because the quantum
mechanical averaging of the wavefunction over the atomic potentials
becomes less systematic in energy regions where the atomic potentials 
differ more.  In this case a single well-defined band persists but
with larger broadening.  In strong scattering systems (which it turns
out this is not), split-band behavior can result. 

Because the behavior near $E_F$ is important for (super)conductivity,
the blowup of the spectral densities near $E_F$ along
the K-$\Gamma$-L line for $x$=0.005 and $x$=0.025 are given in
Fig. \ref{smallband}.
Increasing the B content ($x$) increases the chemical disorder broadening,
in addition to lowering E$_F$.
The change of the disorder broadening at the $\Gamma$ point is
different from that of the L point, because the main B $2p$ character
is located just below $-2$ eV at $x$=0.025. (See below.)
As a result, the broadening at the $\Gamma$ point is half of that 
at the L point at $x=$0.025, while the width at the $\Gamma$ point 
is 10\% wider than at the L point at $x$=0.005.
Thus disorder broadening effects (such as the residual resistivity) 
will be nonlinear in $x$ in this region.

The atom-projected B and C DOS from CPA at $x$=0.025 is 
shown in Fig. \ref{cpados}.
The DOS is similar to those from supercell calculations,
but without sharp peaks in the range of $-2$ eV and 0 eV 
that arise due to the periodic array of dopant B atoms.
The B DOS has a peak at $-2$ eV and decreases asymmetrically.
The C DOS is hardly distinguishable from that of the bulk C.
The weight of the B DOS is closer to E$_F$ than that of the C DOS,
as in the supercells. 
Near $E_F$, the DOS of B is 1.7 times larger than the DOS of C,
which indicates the B influence on metallic properties of C$_{1-x}$B$_x$
is greater than that of an average C atom (but by less than a factor
of two), consistent with measurement by soft X-ray absorption and 
emission spectroscopy.\cite{nakamura2}

\section{Relaxation}
\begin{figure}[tbp]
\rotatebox{-90}{\resizebox{5cm}{8cm}{\includegraphics{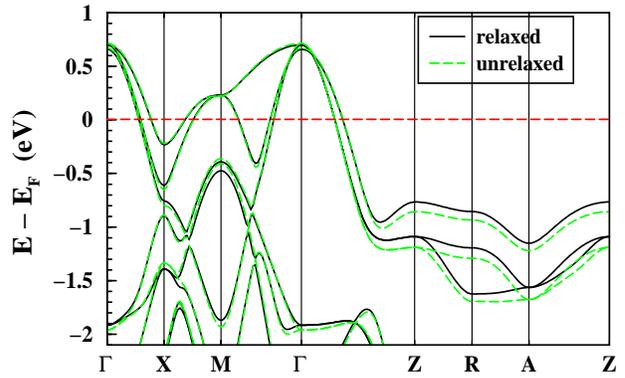}}}
\caption{(Color online) Relaxed and unrelaxed band structures of 
 the B-doped diamond for the $2\times2\times1$ supercell. 
 The relaxation shows little effect. As expected, the strongly hybridized
 bands between the B and nn C on the $k_z=\pi/a$ plane are most 
 affected by the relaxation.
 The difference between the Fermi energies is 25 meV.}
\label{scband}
\end{figure}

Reports of the B-C bond length, when allowed to relax, differ.  The 10\% value
reported in the cluster calculation of Mainwood\cite{mainwood} is clearly
unrealistically large.  Sitch {\it et al.}, using a realistic (but not first principles)
method with large supercell (216 atoms) found a 4\% lengthening,\cite{sitch}
while Blase {\it et al.}, using first principles methods but only a 54 atom
supercell, reported only a 2\% elongation.\cite{blase}  These values are
with respect to 
the C-C bond length of 1.54~\AA~for intrinsic diamond.
The relaxed and unrelaxed band structures of the B-doped diamond for the
$2\times2\times1$ supercell given in Fig. \ref{scband} show that the 
effect of the 2\% elongated B-C bond length is to further raise the 
B-derived bands, {\it i.e.} increase the separation from the bulk C bands,
although the effect is not major (a maximum of 60 meV).
The region between $-1.7$ eV to $-0.7$ eV on the $k_z=\pi/a$ plane 
of this supercell is
most strongest affected, as expected, since that is the
region in which B $2p$ states are strongly mixed with nn C $2p$ states.
This 2\% relaxation results in a small (3\%) increase in the Fermi energy 
(by 25 meV),
which can be considered as a slight increase in the effective mass.
The magnitude of these shifts suggests that relaxation will 
change numerical results by
a few percent but will not change significantly any of the general 
features of B doping that are discussed above.

\section{Summary}
In this paper we have studied the boron spectral density relative to
that of C, and 
quantified to some extent the
disorder broadening, in B-doped diamond, using a combination of 
supercell and random disorder (CPA) approaches.  All experimental 
results that we are aware of support
the prevailing viewpoint that this system is a heavily-doped degenerate
(metallic) semiconductor at the concentration range where superconductivity
has been seen, and our studies lie only within this scenario.
As far as the average distribution of B character in the itinerant
band system is concerned, the CPA and supercell
results are similar.  In the region of interest (around the Fermi level
for reported levels of doping, not exceeding 5\%), the CPA result is
that the fraction of B character is 1.7 times that of a C atom.
Supercell results are consistent with this value, at least if a 
little energy-averaging is considered.  

In the energy-momentum
distribution of B character, however, the periodicity in the supercells
leads to regions where a band has a much larger fraction of B
character than 1.7$x$ (the fraction of B character in B$_xC_{1-x}$
given by the CPA), and is split away from bands that are
primarily C in character.  This effect arises in both the smaller and the
larger supercells (3\% doping down to 0.8\%).  The CPA method, on
the other hand, leads to bands much like those of intrinsic diamond but
disorder broadened to some extent, and with a B content that varies
smoothly along the band (that is, with energy and with momentum).  
The CPA description is buttressed by ARPES data,\cite{arpes}
where mapping of the bands results in diamond-like band topology as
occurs in the VCA or (accounting for disorder) CPA. 

An attractive feature of the supercell method is its ability to
investigate the local environment of the dopant atom, including
lattice relaxations.  While the cluster CPA method makes it 
possible within a disordered system, there is very little experience
with this capability.  The ability to relax atomic positions was  
used by Blase {\it et al.}\cite{blase}, who obtained a 2\% 
longer B-C bondlength than that of diamond; this elongation 
will lead in turn
to a change in the B-C force constant.  Whether this factor 
can account for the difference in spectral distributions obtained
by the two supercell methods\cite{blase,xiang} is yet to be resolved.

\section{Acknowledgments}
We acknowledge K. Koepernik for important technical advice.
This work was supported by National Science Foundation grant No. DMR-0421810.


\begin{thebibliography}{10}
\bibitem{2nnC} {In the unit cell, there are two types of 2nd n C. 
 One is surrounded by two B on the a-b plane, and another is along 
 the diagonal direction from B.
 While the former is affected more strongly by the B, the latter acts 
 like the bulk C. So we call only the former as 2nd n C in this paper.}
\end{thebibliography}
\end{document}